\documentclass[prl,nofootinbib,twocolumn,showpacs]{revtex4}
\usepackage{graphicx}%
\usepackage{dcolumn}
\date{today}
\def\bq{\begin{equation}}
\def\ee{\end{equation}}
\begin{document}
\begin{abstract}
The  Landauer transport formulation is generalized to the case of
a dynamic scatterer with an arbitrary energy level structure,
 weakly coupled to a long ideal noninteracting wire.
 The two-terminal {\em linear}
conductance of the device is expressed in terms of the (in general
inelastic) scattering cross sections of the electrons off the
scatterer, using a Fermi-liquid picture for the scattering states
of the whole system. Assuming unitarity, i.e. the optical theorem,
we obtain the same results from a Kubo linear response treatment
and from a Keldysh technique approach.
\end{abstract}
\title{A Landauer transport formulation with inelastic scattering}
\author{Yoseph Imry$^{1}$, Ora Entin-Wohlman$^{2,3,4}$ and Amnon
Aharony$^{2,3}$} \affiliation{$^{1}$ Dept. of Condensed-Matter
Physics, the Weizmann Institute of Science, Rehovot 76100,
Israel\\ $^{2}$ School of Physics and Astronomy, Raymond and
Beverly Sackler Faculty of Exact Sciences,  Tel Aviv University,
Tel Aviv 69978, Israel\\ $^{3}$ Physics Department, Ben Gurion
University, Beer Sheva 84105,
Israel\\
$^{4}$ The Albert Einstein Minerva Center for Theoretical
Physics\\ at the Weizmann Institute of Science, Rehovot 76100,
Israel} 
\date{\today}
\maketitle


\noindent {\bf Introduction.} The scattering model due to Landauer
\cite{Land} has   been extremely useful for studies of the
electronic transport in mesoscopic systems, since it gives a clear
physical picture of the processes involved (for reviews, see Refs.
\onlinecite{IL} and \onlinecite{book}). This formulation expresses
the linear conductance ${\cal G}$ of a given finite sample (the
scatterer) in terms of transmission probabilities for electrons
injected to and collected from it, via ideally conducting leads.
For noninteracting electrons, with a total transmission amplitude
$t$ and magnitude ${\cal T}=|t|^2$, the two-terminal
single-channel Landauer formula, including spin, reads
\begin{equation} {\cal G}=\frac{2e^2}{h}{\cal T}.
\label{land} \end{equation} In this paper we generalize his result
to include inelastic scattering by the sample.

A general formal expression for the conductance of a scatterer
imbedded in a long, ideal, ``wire" was given in Ref. \onlinecite
{MW}. It was proven to reduce to the Landauer formula (\ref{land})
at the  zero temperature ($T \rightarrow 0$) limit. However, at
finite $T$ it was not possible to produce such a proof, and the
inelastic scattering channels, which open up then, were
implicated. Later, the opinion that the Landauer formula is not
applicable with interactions on the scatterer became prevalent
(e.g. Ref. \onlinecite{KG}). That would mean that many of the
usual mesoscopic effects would be severely modified by the
inelastic scattering and interactions. Thus, the generalization
(see also Ref. \onlinecite{AT}) we propose is relevant. Indeed,
the {\em full} scattering matrix of  a local perturbation in an
otherwise ideal conducting wire, should contain  the necessary
physical information for the conductance. The quasiparticle
picture is necessary  to approximately include the effects of
electron-electron interactions on the leads mediated by virtual
second-order processes on the scatterer. Within this picture,
multiparticle processes and non-Fermi-liquid behavior are avoided.

The linear dc  conductance of the sample, at finite temperatures,
is simply expressed  through its full single-quasiparticle
scattering matrix (satisfying unitarity and time-reversal
symmetries). The results are then confirmed via the Kubo linear
response approach \cite{Kubo}, considering the sample as a set of
scattering states with  equilibrium populations. Finally,  we use
the optical theorem \cite{glazman}  to express the linear Keldysh
\cite{jauho,wingreen,datta,MW} conductance in terms of the
sample's transmission cross-sections, also confirming the
generalized Landauer result.

\noindent {\bf Scattering formulation for a one-dimensional (1D)
conductor.} We demonstrate our statement for the simple case of a
scatterer (a ``quantum dot"), taken as a {\it finite} ``black box"
with {\it arbitrary interactions inside it}, coupled weakly to
ideal 1D leads. We start by writing the scattering states,
including inelastic processes. The states coming from the left
($\ell$)  and from the right ($r$) belong to two independent
Hilbert spaces, with no mutual Fermi blocking factors.  For linear
transport, the two chemical potentials of the ($\ell$) and ($r$)
states, $\mu_\ell$ and $\mu_r$ respectively, are slightly
different ($\mu_r < \mu_\ell$, $\mu_\ell - \mu_r \equiv eV <<
\mu_{\ell, r}$). At equilibrium, both spaces are populated with
the same temperature $T$ and $\mu_\ell=\mu_r=\mu$. Without the
coupling, the states $|i \rangle$ of the dot have discrete energy
levels, $\epsilon_i$, including the arbitrary interactions there.
The leads have  continuous spectra, characterized by a wavenumber
$k$, longitudinal velocity $v \equiv \hbar k /m$ and a kinetic
energy of the ``incoming" electron $ E \equiv \hbar^2 k^2/2m$. The
states coming from the left lead (labelled by $\ell$) are
\vspace{-0.05cm}
\begin{eqnarray} \nonumber \mid \ell k i\rangle = e^{ikx} \mid i\rangle
+ \sum_j r_{ji} \sqrt{\frac{v}{v_{j}}}\, e^{-ik_{j} x} \mid
j\rangle ,\\{\rm and}~~~~ \sum_j t_{ji}
\sqrt{\frac{v}{v_{j}}}\,e^{ik_{j} x} \mid j\rangle , \label{l}
\end{eqnarray}
\noindent far to the left of the dot and
far on its right, $v_j=\hbar k_{j}/m~(>0)$ is defined  by $
\hbar^2 k_{j}^2/2m \equiv E - \epsilon_j + \epsilon_i $, $t_{ji}$
and $r_{ji}$ are the transmission and reflection amplitudes from
the left, exciting the dot from state $|i\rangle$ to state
$|j\rangle$. The states coming from the right lead are
\begin{eqnarray}
\nonumber \mid rki\rangle = \sum_j t'_{ji }
\sqrt{\frac{v}{v_{j}}}\, e^{-ik_jx} \mid j\rangle, ~~~~{\rm and}\\
e^{-ikx} \mid i\rangle + \sum_j r'_{ji} \sqrt{\frac{v}{v_{j}}}\,
e^{ik_{j} x} \mid j\rangle, \label{r}\end{eqnarray} \noindent far
to the left of the scatterer and far on its right, with the
transmission and reflection amplitudes $t'_{ji}$ and $r'_{ji}$.
Using ${\cal T}_{ij} \equiv |t_{ij}|^2,~ {\cal T}'_{ij} \equiv
|t'_{ij}|^2,~ {\cal R}_{ij} \equiv |r_{ij}|^2, ~{\cal R}'_{ij}
\equiv |r'_{ij}|^2$, the total right-going current in the states
(\ref{l}) with an energy $E$ in a window $dE$, per spin direction,
is
\begin{equation}
 I_{\ell ki} = e v \sum_j {\cal T}_{ji} n_i dE = \frac{e
}{2\pi \hbar } \sum_j {\cal T}_{ji}  dE, \label{zerem}
\end{equation}
where $n_i = (2\pi \hbar v)^{-1}$ is the density of $\{\ell ki\}$
states per spin channel (which cancels the velocity factor, as
usual).

We next discuss the thermal occupations.  The probability to find
the dot in state $|i \rangle$ is $P_i$, with $\sum_i P_i=1$.
Consider a typical low-lying excited state of the whole system
with the dot in state $|i \rangle$. Within a Fermi liquid
description the number of quasiparticles in this state is
macroscopic. Thus the contribution of the dot energy $\epsilon_i$
to the energy of a single quasiparticle state is negligible. Hence
their populations are given by the Fermi function {\it of the
electronic energy only}. Thus, the populations of the scattering
states $\mid \ell ki\rangle$ and $\mid rki\rangle$ are
\vspace{-0.05cm} \bq P_{\ell ki, r ki} = P_i f(E  - \mu_{\ell,
r}), \label{eqpop}\ee \noindent where $f(x)=1/(e^{\beta x}+1)$ is
the Fermi function and $\beta = (k_{B} T)^{-1}$. The total
populations of incoming electrons with energy $E$ are given by
$f_{\ell,r}(E)=f(E  - \mu_{\ell, r})$.

The averaged right (left)-going current carried  by the $\ell$
($r$) states is given by \begin{eqnarray} \nonumber I_{\ell
\rightarrow r} &=& \sum_iP_{\ell ki}I_{\ell ki}=\frac{e}{2\pi
\hbar}
\int dE \sum_{ij}P_{\ell ki} {\cal }T_{ji}; \label{Ilr}\\
\noindent I_{r\rightarrow \ell} &=& \frac{e}{2\pi \hbar} \int dE
\sum_{ij}P_{rki} {\cal T}'_{ji}.\label{Irl}\end{eqnarray}
\noindent For a finite,  time-reversal
symmetric dot, with real eigenstates,
${\cal T}_{ji}= {\cal T}'_{ji},$
meaning that the cross-section for scattering of an electron from
the dot, flipping the dot from state $i$ to state $j$, has the
same value irrespective of the direction from which the electron
is coming. The total average net current is given by $I = I_{\ell
\rightarrow r} - I_{r\rightarrow \ell}$. Using Eq. (\ref{eqpop}),
we find, to first order in  $eV \equiv \mu_\ell - \mu_r$, that the
linear conductance, ${\cal G}\equiv \lim_{V \rightarrow 0} I / V$
per spin is \bq {\cal G}  = \frac{e^2} {2\pi \hbar } \int dE \Bigl
( -\frac{\partial f}{\partial E}\Bigr )_{\mu} \sum_{ij} P_i {\cal
T}_{ji} .\label{result}\ee

\noindent This is the two-terminal single-channel,
finite-$T$ Landauer linear conductance formula
including  arbitrary interactions on the scatterer and inelastic
scattering.\\
\noindent{\bf Rederivation from the Kubo formula.} Here, we
confirm our main result, Eq. (\ref{result}), using the $\omega
\rightarrow 0$ limit of the Kubo linear response formulation
\cite{Kubo},
\begin{eqnarray}
{\cal G}(\omega)={  \pi
e^2\over{
 \omega}}[S_{v}(-\omega) - S_{v}(\omega)].
\label{fdt}
\end{eqnarray}
\noindent The velocity power spectrum, $S_{v}(\omega)$ is given
for  Fermi liquid electrons in a unit volume  and including spin
by
\begin{eqnarray} \nonumber
S_{v}(\omega) \equiv \sum_{m,n} |\hat v_{m n} |^2 \delta(\hbar
\omega- E_m + E_{n} )\times \\  P_i f(E_{m} - \mu) [1 - f(E_{n}  -
\mu )]\, , \label{Kubo-e}\end{eqnarray}
%
\noindent where the matrix elements of the velocity operator $\hat
v$ (in the wire direction) are taken between an initial state  $|m
\rangle = |\ell ki \rangle$, or $|rki \rangle$ and a final one,
$|n \rangle = |\ell k' i' \rangle$, or $|rk' i' \rangle$.  $k$ is
related to $E = E_m$ and $k'$ to $E' = E_n$. The summation over
$m$ means summation over $i$, taking the sum of the $\ell$ and the
$r$ terms and integrating over the energy $E$ (which, as explained
above, does not contain the dot energy), with the appropriate
density of states, and similarly for the summation over $n$. Note
that here the Fermi blocking factors, $[1 - f(E_{n} - \mu )]$,
{\it do} appear, since {\em real transitions} are considered. The
conductance is proportional to the {\it net} absorption  rate from
a classical field at frequency $\omega$  \cite{uri}.

The transitions are of four types: $(\ell,\ell)$ (left-moving to
left-moving), $(\ell,r)$ (right-moving to left-moving), and
similarly $(r,r)$ and $(r,\ell)$. The sum of the two latter
transitions equals that for the two former ones, so we calculate
only the former, and multiply the final result by two.  From Eq.
(\ref{l}), \vspace{-0.05cm}
\begin{eqnarray} \langle \ell ki| \hat v |\ell k'i' \rangle
& = & \frac{ \sqrt{v v'}}{2} \sum_j ( t_{ji}^\ast t_{ji'} +
\delta_{i, i'} - r_{ji}^\ast r_{ji'}) \nonumber
\\ & = &
\sqrt{v v'} \sum_j t_{ji}^\ast t_{ji'} ,
\end{eqnarray}
\noindent where $v' = \hbar k' /m$ and we used the (column)
unitarity properties of the full S-matrix, $ \sum_j t_{ji}^\ast
t_{ji'}   = \delta_{i, i'} - \sum_j r_{ji}^\ast r_{ji'}$.
Likewise, $\langle \ell ki| \hat v |rk'i' \rangle = \sqrt{v v'}
\sum_j t_{ji}^\ast r'_{ji'}.$ Taking the absolute values squared
of these matrix elements, summing over $i'$, (together with the
density-of-states factors required to convert the  $k$- and
$k'$-sums into energy integrals) and using the (row) unitarity
properties of the full S-matrix, $\sum_{i'}( t^\ast_{ji'} t_{j'i'}
+ r'^\ast_{ji'} r'_{j'i'})  = \delta_{j, j'}$, yields \bq
\sum_{i'}n_i n_{i'}\bigl (|\langle \ell ki| \hat v | \ell k'i'
\rangle |^2 + |\langle \ell ki| \hat v | rk'i' \rangle |^2 \bigr
)= \frac{1}{h^2}\sum_{j} {\cal T}_{ji}. \label{me} \ee Adding a
similar contribution from the states which start in $|rki
\rangle$, we put the resulting expression into
Eq. (\ref{Kubo-e}), and use $f(E + \hbar \omega/2)[1- f(E - \hbar
\omega/2)] = f(E )[1- f(E)] + (\hbar \omega/2) f'(E)+{\cal
O}(\omega^2)$, to obtain for  small $\omega$,
\begin{eqnarray} \nonumber
S_{v}(\omega) =\frac{2}{ h^{2}}  \int dE \bigl [f (1 -f) + (\hbar
\omega/2)\Bigl (\frac{\partial f}{\partial E}\Bigr )\bigr
]\sum_{ij}
 P_i {\cal T}_{ji}.\label{S-v}\end{eqnarray}
\noindent  Eq. (\ref{fdt}) now yields Eq.
(\ref{result}) for ${\cal G}(\omega=0)$, as required.

\noindent{\bf Confirmation via the Keldysh technique.}
It has been claimed \cite{MW,KG} that when the transmitted
electron is subject to inelastic interactions on the scatterer,
the transmitted current should be obtained using methods
applicable to many-body problems. The most-frequently used one is
the Keldysh technique. Here we calculate independently the
transmission of the system from the full scattering operator, and
the current using the Keldysh technique, and then confirm that the
two expressions so obtained  are identical. Moreover, this
equivalence is just a result of the {\it optical theorem}, namely,
it follows from the assumed unitarity of the {\em full} scattering
matrix.

Considering a quantum dot with a single electronic level, the
coupling with the leads  is $\sum_kV_{\ell k}c_{\ell k}^\dagger
d+\sum_pV_{r p}c_{r p}^\dagger d +hc$. (The band states on the
left are denoted by the wavevector $k$ and those on the right by
$p$. Both leads are identical, except of being connected to
reservoirs of different chemical potentials. $d^\dagger$ creates
an electron on the dot.) The current between the leads is given by
the one flowing from the left, $I_{\ell}=e(d/dt)\sum_k \langle
c_{\ell k}^\dagger c_{\ell k} \rangle$ (where $\langle ...
\rangle$ is the quantum average), or by its counterpart $-I_r$. In
the Keldysh technique these currents are expressed in terms of
various Green functions, resulting in (see, e.g., Ref.
\onlinecite{jauho})
\begin{eqnarray}
I=\frac{e}{\hbar}\int d\omega \Bigl (f_{\ell}-f_{r}\Bigr )\Bigl
[\frac{i}{\pi}\frac{\Gamma_{\ell}\Gamma_{r}}{\Gamma_{\ell}+\Gamma_{r}}\Bigl
(\overline{G^{R}_{d}}-\overline{G^{A}_{d}}\Bigr )\Bigr
],\label{keldysh} \end{eqnarray} where the partial widths,
\begin{eqnarray}
\Gamma_{\ell ,r}(\omega )&=&\pi\sum_{k}|V_{\ell k,rp}|^{2}\delta
(\hbar\omega
-\epsilon_{k,p}),
\label{gamma}
\end{eqnarray}
are needed only in a narrow energy interval around the (average)
Fermi level of the leads. Therefore, one may neglect the
energy-dependence of these widths \cite{glazman}.
$\overline{G^{R,A}_{d}}(\omega)$
are the thermal averages of the usual retarded and advanced Green
functions on the dot, which include all interactions. Despite the
apparent similarity of Eq. (\ref{keldysh}) to the Landauer
formula, it has been claimed \cite{jauho} that there is no
connection between the terms in the square brackets there and the
transmission, once inelastic scattering is present. However, as we
show below, these terms are in fact identical to the transmission.

The transmission is derived from the full scattering operator of
the system, $S$.  Writing  the total Hamiltonian of the system as
${\cal H}={\cal H}_{0}+{\cal H}_{1}$, where ${\cal H}_{0}$ is the
Hamiltonian of the disconnected system (including the interactions
on the dot) and ${\cal H}_{1}$ describes the coupling between the
dot and the leads, then (to all orders in ${\cal H}_1$), $S=\lim_{
t\rightarrow\infty}U(t)$, where
\begin{eqnarray}
&&U(t)=1-\frac{i}{\hbar}\int_{-\infty}^{t}dt'e^{i\frac{{\cal
H}_{0}}{\hbar}t'}{\cal H}_{1}e^{-i\frac{{\cal
H}_{0}}{\hbar}t'}\nonumber\\
&-&\int_{-\infty}^{t}\frac{dt'}{\hbar}\int_{-\infty}^{t'}\frac{dt''}{\hbar}
 e^{i\frac{{\cal H}_{0}}{\hbar}t'}{\cal
H}_{1}e^{-i\frac{{\cal H}}{\hbar}(t'-t'')}{\cal
H}_{1}e^{-i\frac{{\cal H}_{0}}{\hbar}t''}.\nonumber\\
\label{S}
\end{eqnarray}
The transition probability per unit time between two states of the
system, $|m\rangle$ and $|n\rangle$, is given by
\begin{eqnarray}
W_{m\rightarrow n}=\lim_{t\rightarrow\infty}\Bigl (\langle
m|U^{\dagger}|n\rangle\frac{d}{dt}\langle n|U|m\rangle +cc\Bigr ).
\label{W}
\end{eqnarray}
When one specifies to the case in which both states $|m\rangle$
and $|n\rangle$  have no electron on the dot and differ only by
one electron being on the left lead in the former and on the right
lead in the latter, one finds \cite{glazman,wingreen}
\begin{eqnarray}
&&W_{k\rightarrow p}=\frac{|V_{\ell k}V_{r
p}|^{2}}{\hbar^{4}}\int_{0}^{\infty}d\tau_{1} d\tau_{2} d\tau_{3}
e^{i\frac{\epsilon_{k}(\tau_{3}-\tau_{1})}{\hbar}}
e^{i\frac{(\epsilon_{p}-\epsilon_{k})\tau_{2}}{\hbar}}\nonumber\\
&&\times G_{d}^{A}(-\tau_{2}-\tau_{1},
-\tau_{2})G_{d}^{R}(0,-\tau_{3}) +cc.\label{Wkp}
\end{eqnarray}
The Green functions appearing here depend on two time arguments
(and not on their difference alone), since the thermodynamic
average has not yet been performed.
It is of crucial
importance to realize that the Green functions appearing in the
Keldysh technique result (\ref{keldysh}) for the (linear response)
current
are the Fourier transform of the
{\it thermodynamically averaged} Green functions. On the other
hand, the transition probability per unit time
includes a product of {\it two} Green functions. Obviously, the
thermodynamic average of the product is not necessarily the
product of the two averages. The relation between the two is
supplied by the optical theorem.

The transmission from a state of an electron with energy
$E=\hbar\omega$ on the left lead to a state of an electron with
energy $E'=\hbar\omega '$ on the right lead is \cite{glazman}
${\cal T}(\omega\rightarrow\omega ')=\sum_{
k,p}\overline{W}_{k\rightarrow p}\delta
(\frac{\epsilon_{k}}{\hbar}-\omega)\delta
(\frac{\epsilon_{p}}{\hbar}-\omega ').$ Hence, the total
transmission probability of an electron having initial energy
$\hbar\omega $ is
${\cal T}(\omega )=\int d\omega '{\cal T}(\omega\rightarrow\omega
').$
For  overall energies near $\mu$, ${\cal T}(\omega )$ is identical
to the quantity $  \sum_{ij}  P_i{\cal T}_{ji}$ which appeared in
Eq. (\ref{Ilr}). With the definitions (\ref{gamma}), ${\cal
T}(\omega )$ becomes
\begin{eqnarray}
{\cal T}(\omega )&=&\frac{2\Gamma_{\ell}\Gamma_{r}}{\hbar^{2}\pi}
\int_{0}^{\infty }d\tau_{1}d\tau_{2}
e^{i\omega
(\tau_{2}-\tau_{1})}\nonumber\\
&&\times\overline{G_{d}^{A}(-\tau_{1},0)G_{d}^{R}(0,-\tau_{2})},\label{T}
\end{eqnarray}
and consequently, using the Landauer formulation (see also Ref.
\onlinecite{datta})
\begin{eqnarray}
I=e \int d\omega \Bigl (f_{\ell}(\omega )-f_{r}(\omega )\Bigr
){\cal T}(\omega ).\label{landauer}
\end{eqnarray}

Next  we show that the two expressions, Eqs. (\ref{keldysh}) and
(\ref{landauer}), are identical, namely, that the terms in the
square brackets in Eq. (\ref{keldysh}) are given by ${\cal
T}(\omega )$, Eq. (\ref{T}). To this end, we use Eq. (\ref{S}) to
write down explicitly the condition that $S$ is unitary,
$\sum_{m}\langle n|S^{\dagger}|m\rangle\langle m|S|l\rangle
=\delta_{n,l}$. Specifying  to the case in which  the states
$|l\rangle$ and $|n\rangle $ differ just by one electron taken
from the left to the right lead, the unitarity condition becomes
\begin{eqnarray}
&&0=\frac{V_{\ell k}^{\ast}V_{r
p}}{\hbar^{4}}\int_{-\infty}^{\infty}dt'
dt''e^{i(\frac{\epsilon_{p}}{\hbar}t'-\frac{\epsilon_{k}}{\hbar}t'')}\Bigl
[i\Bigl (G^{A}_{d}(t',t'')\nonumber\\
&-&G^{R}_{d}(t',t'')\Bigr ) +
\int_{-\infty}^{\infty}dt_{1}dt_{2}G^{A}_{d}(t',t_{1})G^{R}_{d}(t_{2},t'')\nonumber\\
&& \times \Bigl (\sum_{p'}|V_{r
p'}|^{2}e^{i\frac{\epsilon_{p'}(t_{2}-t_{1})}{\hbar}}+\sum_{k'}
|V_{\ell
k'}|^{2}e^{i\frac{\epsilon_{k'}(t_{2}-t_{1})}{\hbar}}\Bigr )
\Bigr] .\nonumber\\
\label{optical}
\end{eqnarray}
Since $\Gamma_{\ell ,r}$ are independent of the energy, we have
$\sum_{k,p}|V_{\ell
k,rp}|^{2}e^{i\frac{\epsilon_{k,p}}{\hbar}t}=2\Gamma_{\ell,r}\delta
(t).$
Performing   the thermodynamic average over Eq. (\ref{optical}),
taking into account  that $G^{R}(t_{1},t_{2})$
($G^{A}(t_{1},t_{2})$) is defined for $t_{1}\geq t_{2}$
($t_{1}\leq t_{2}$), and that $\overline{G^{R,A}(t_1,t_2)}$
depends only on $(t_1-t_2)$, we find
\begin{eqnarray}
&-&iV_{\ell k}^{\ast}V_{r p}\Bigl [2\pi\delta
(\epsilon_{k}-\epsilon_{p})\Bigl
(\overline{G^{R}_{d}(\epsilon_{k})}-
\overline{G^{A}_{d}(\epsilon_{k})}\Bigr )\nonumber\\
&+&2i\frac{\Gamma_{\ell}+\Gamma_{r}}{\hbar^{2}}\int_{-\infty}^{\infty}dt
e^{i\frac{(\epsilon_{p}-\epsilon_{k})t}{\hbar}}
\int_{0}^{\infty}d\tau_{1}d\tau_{2}
e^{i\frac{\epsilon_{k}(\tau_{2}-\tau_{1})}{\hbar}}\nonumber\\
&&\times\overline{G^{A}_{d}(t-\tau_{1},t)G^{R}_{d}(t,t-\tau_{2})}\Bigr
]=0.\label{optical1}
\end{eqnarray}
Since the thermal average in the second term here is independent
of $t$, the $t-$integration  yields $2\pi\delta
(\epsilon_{k}-\epsilon_{p})$ and Eq. (\ref{optical1}) gives
\begin{eqnarray}
\overline{G^{R}_{d}(\epsilon_{k})}-\overline{G^{A}_{d}(\epsilon_{k})}=-
i\pi\frac{\Gamma_{\ell}+\Gamma_{r}}{\Gamma_{\ell}\Gamma_{r}}{\cal
T}(\epsilon_{k}), \label{equality}
\end{eqnarray}
which shows that the square brackets in Eq. (\ref{keldysh}) are
equal to ${\cal T}(\omega )$, Eq. (\ref{T}). That is, the current
using the Keldysh method, Eq. (\ref{keldysh}), and the same
current by the Landauer-type formula, Eq. (\ref{landauer}), are
identical.  We have confirmed  this equivalence explicitly by
considering the case in which the electrons on the dot interact
with a phonon bath \cite{EAIfuture}.


\noindent {\bf Conclusions.} Within the Landau  framework (used
here for single-quasiparticle scattering states), we found that
the Landauer-type picture for the {\it linear} transport is
generalizable to include {\it inelastic} scattering and
interactions on the ``dot". We considered here a dot having, when
it is closed, an arbitrary energy level structure, due to
(possibly strongly interacting)  degrees of freedom (e.g.
vibrations, spins) {\it other} than those of the conduction
electrons. The results were confirmed with both the linear
response and Keldysh approaches. Unitarity and time-reversal
symmetries for the dot played crucial roles. However, the
generalization for a dot imbedded in an Aharonov-Bohm
interferometer is straightforward. So are those to multichannel
and multiterminal situations.

\noindent {\bf Acknowledgements.} This work was
supported by  a Center of Excellence of the Israel Science
Foundation
and by the German Federal Ministry of Education and Research
(BMBF), within the framework of the German Israeli Project
Cooperation (DIP). It was stimulated by discussions that took
place within the Nanoscience program at the Institute for
Theoretical Physics, the University of California, Santa Barbara
in the fall of 2001. Discussions with Y. Alhassid, N. Andrei, J.
von Delft, Y. Levinson, Y. Meir, P. Mello, Z. Ovadyahu, A. Rosch,
B. Shapiro, U. Smilansky, H. Weidenm\"uller, N. Wingreen and
especially with the late R. Landauer and with U. Gavish, are
gratefully acknowledged.

\end{document}